\documentclass{egpubl}
\usepackage{eg2027}
\pdfSubject{Preprint}
\copyrightTextTitPag{}
\copyrightTextRunPag{}
\makeatletter
\let\ps@titlepage\ps@plain
\makeatother
\AtBeginDocument{\hypersetup{pdfkeywords={},pdfcreator={},pdfproducer={}}}

\usepackage[T1]{fontenc}
\usepackage{dfadobe}  
\usepackage{enumitem}
\usepackage{cite}  %
\BibtexOrBiblatex
\electronicVersion
\PrintedOrElectronic
\ifpdf \usepackage[pdftex]{graphicx} \pdfcompresslevel=9
\else \usepackage[dvips]{graphicx} \fi

\usepackage{algorithm}
\usepackage{tabularx}
\usepackage{algpseudocode}
\usepackage{booktabs}

\usepackage{egweblnk}

\usepackage{math-notation}   
\usepackage{paper-macros}    
\usepackage{xspace}
\usepackage[table]{xcolor}

\newcommand{\methodname}{Deformable Texture Maps}
\newcommand{\methodabbr}{DTM\xspace}

\newcommand{\neuraltexture}{\mathcal{T}}

\newcommand{\mat}[1]{\mathbf{#1}}

\usepackage[font=it,labelfont={bf,up},labelsep=colon,skip=6pt,hypcap=false]{caption}
\usepackage[capitalise,noabbrev]{cleveref}

\usepackage[strings]{underscore}

\providecommand{\citep}{\cite}

\title[\methodname: Compiling Dynamic Microgeometry into Neural Textures]{\methodname: Compiling Dynamic Microgeometry into Neural Textures}

\author[A. A. Silaghi et al.]{%
  \parbox{\dimexpr\textwidth-2pt\relax}{\centering
    \href{mailto:antonioarian@yahoo.es}{Antonio Arian Silaghi}$^1$,
    \href{mailto:zhenjaychen@gmail.com}{Zhen Chen}$^2$,
    \href{mailto:dannykaufman@gmail.com}{Danny Kaufman}$^2$,
    \href{mailto:emichel@adobe.com}{\'{E}lie Michel}$^2$,\\
    \href{mailto:jeanmarc.thiery@gmail.com}{Jean-Marc Thiery}$^2$,
    \href{mailto:jorge@jorg3.com}{Jorge Lopez-Moreno}$^1$,
    \href{mailto:elenagarces@gmail.com}{Elena Garces}$^2$
  }\\
  \parbox{\dimexpr\textwidth-2pt\relax}{\centering
    $^1$University Rey Juan Carlos\qquad $^2$Adobe
  }
}

\begin{document}

\teaser{
 \captionsetup{skip=3.46pt,belowskip=0pt}
 \includegraphics[width=0.9\linewidth]{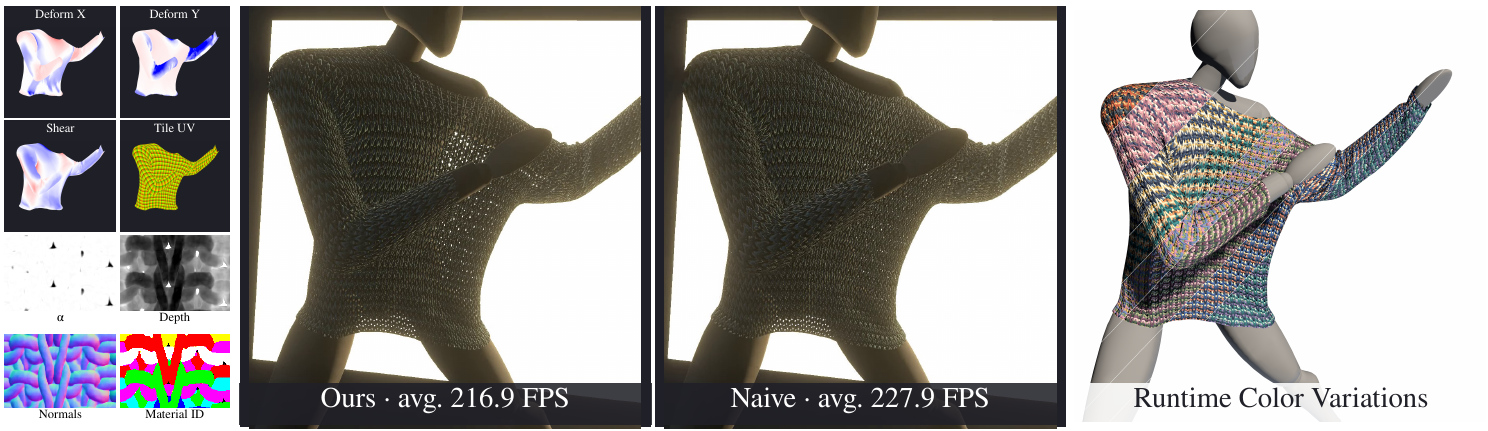}
 \centering
\caption{
We provide high-resolution neural textures that are conditioned on the base mesh stretch values. This allows representing non-trivial local variations of the texture with a single parametric model, which are required to capture deformation behaviors of cloths under deformation (compare "Ours" with "Naive", which corresponds to simple texture stretch). Our textures remain parametric, and color variations of the various yarns composing the cloth texture can be explored at run-time (right image).
}
\label{fig:teaser}
}

\maketitle

\defresultval{sweep_h128_1em3}{20.10}
\defresultval{sweep_h128_1em4}{20.70}
\defresultval{sweep_h128_3em3}{11.80}
\defresultval{sweep_h128_3em4}{21.08}
\defresultval{sweep_h128_alpha}{20.53}
\defresultval{sweep_h128_depth}{22.23}
\defresultval{sweep_h128_normals}{20.48}
\defresultval{sweep_h256_1em3}{15.18}
\defresultval{sweep_h256_1em4}{22.60}
\defresultval{sweep_h256_3em3}{11.80}
\defresultval{sweep_h256_3em4}{22.81}
\defresultval{sweep_h256_alpha}{22.28}
\defresultval{sweep_h256_depth}{23.65}
\defresultval{sweep_h256_normals}{22.49}
\defresultval{sweep_h32_1em3}{16.51}
\defresultval{sweep_h32_1em4}{16.25}
\defresultval{sweep_h32_3em3}{16.45}
\defresultval{sweep_h32_3em4}{16.52}
\defresultval{sweep_h32_alpha}{15.80}
\defresultval{sweep_h32_depth}{18.40}
\defresultval{sweep_h32_normals}{15.35}
\defresultval{sweep_h64_1em3}{18.82}
\defresultval{sweep_h64_1em4}{18.41}
\defresultval{sweep_h64_3em3}{11.80}
\defresultval{sweep_h64_3em4}{18.90}
\defresultval{sweep_h64_alpha}{18.23}
\defresultval{sweep_h64_depth}{20.34}
\defresultval{sweep_h64_normals}{18.14}
\defresultval{table4_astc_12_12_alpha}{18.42}
\defresultval{table4_astc_12_12_depth}{24.28}
\defresultval{table4_astc_12_12_group_id}{94.67}
\defresultval{table4_astc_12_12_normals}{18.68}
\defresultval{table4_astc_12_12_storage}{11.02}
\defresultval{table4_astc_4_4_alpha}{34.03}
\defresultval{table4_astc_4_4_depth}{43.54}
\defresultval{table4_astc_4_4_group_id}{99.99}
\defresultval{table4_astc_4_4_normals}{30.75}
\defresultval{table4_astc_4_4_storage}{97.52}
\defresultval{table4_astc_6_6_alpha}{26.44}
\defresultval{table4_astc_6_6_depth}{34.58}
\defresultval{table4_astc_6_6_group_id}{99.79}
\defresultval{table4_astc_6_6_normals}{24.74}
\defresultval{table4_astc_6_6_storage}{43.02}
\defresultval{table4_astc_8_8_alpha}{22.63}
\defresultval{table4_astc_8_8_depth}{28.73}
\defresultval{table4_astc_8_8_group_id}{98.76}
\defresultval{table4_astc_8_8_normals}{21.65}
\defresultval{table4_astc_8_8_storage}{24.77}
\defresultval{table4_neural_h128_alpha}{20.88}
\defresultval{table4_neural_h128_depth}{29.68}
\defresultval{table4_neural_h128_group_id}{96.18}
\defresultval{table4_neural_h128_normals}{21.03}
\defresultval{table4_neural_h128_storage}{0.263}
\defresultval{table4_neural_h256_alpha}{22.60}
\defresultval{table4_neural_h256_depth}{32.86}
\defresultval{table4_neural_h256_group_id}{97.45}
\defresultval{table4_neural_h256_normals}{23.49}
\defresultval{table4_neural_h256_storage}{0.882}
\defresultval{table4_neural_h32_alpha}{15.97}
\defresultval{table4_neural_h32_depth}{24.00}
\defresultval{table4_neural_h32_group_id}{88.18}
\defresultval{table4_neural_h32_normals}{15.20}
\defresultval{table4_neural_h32_storage}{0.044}
\defresultval{table4_neural_h64_alpha}{18.45}
\defresultval{table4_neural_h64_depth}{26.46}
\defresultval{table4_neural_h64_group_id}{93.08}
\defresultval{table4_neural_h64_normals}{18.38}
\defresultval{table4_neural_h64_storage}{0.093}
\defresultval{table4_png_mib}{187.11}
\defresultval{table4_raw_mib}{387.00}

\begin{abstract}

Deformable material surfaces, such as cloth, owe their appearance to fine-scale geometries that restructure under deformation. Even when coarse simulation models efficiently recover this microstructure, explicitly instantiating and rendering it remains expensive.
We introduce Deformable Texture Maps (DTM), a representation that compiles deformation-driven microgeometry into a continuous and compact neural texture. 
Our model maps texture coordinates and local deformations to intrinsic maps describing opacity, surface normals, depth, and material identity. This captures the occlusion and visibility changes generated by deformation. 
At runtime, 
these maps provide a common texture-space representation of deformation-driven fine-scale structure that can be evaluated directly on coarse deforming surfaces.
Since our representation encodes geometric attributes rather than radiance, lighting and material response remain decoupled from the learned model. This allows us to apply relighting, as well as shading and material editing without retraining.
We further introduce a persistent texture-space rendering strategy that caches these intrinsic maps and selectively updates them as deformation changes. This effectively decouples neural evaluation from frame rendering and enables their reuse across frames and conventional rendering effects.
We demonstrate our approach on yarn-based textiles, where non-linear material stretch and shear induce apply changes in yarn geometry spacing and occlusion. We build our model by, first, sweeping tileable yarn patches through deformation space, then, rendering the resulting detailed geometry into a canonical texture-space, and finally, compressing the resulting maps into a single neural representation. 
Our resulting model can be evaluated directly on arbitrary coarse garment meshes and integrated with conventional shading pipelines. 
Our results demonstrate a compact and flexible representation of deformation-driven fine-scale structure while keeping appearance authorable at render time.

\end{abstract}

\section{Introduction}
\label{sec:introduction}

Materials with complex fine-scale structure---threads, fibers, scales, pores---derive their detailed appearance from fine-scale geometric details that are expensive to represent and render explicitly. As they are deformable materials, this structure then evolves with stretching, compression  and shear, to generate or close gaps, changing surface orientations. 
Faithfully reproducing these effects is challenging as it requires capturing fine-scale geometries under deformation.

Textiles are a critical and challenging instance of this problem. The appearance of woven and knitted fabrics emerges from large numbers of interlocking yarns whose spatial arrangement changes as the cloth deforms. 
Yarn-level methods capture this structure explicitly and can further reconstruct fiber-scale detail for rendering~\cite{montazeri2019mechanics,yuksel2012stitchmeshes}, while homogenized models recover detailed yarn configurations from the deformation of a coarse cloth surface modeled with shells~\cite{sperl2020hylc,sperl2021madypg}.
Recent methods further demonstrate that deformation-dependent yarn geometries can be reconstructed and rendered at interactive rates~\cite{huang2025real}
(30ms/frame, fullHD). However, these approaches retain the fine-scale geometry as part of the runtime representation, coupling reconstruction and rendering to the represented microgeometry.
Surface-based fabric appearance models take the opposite approach, avoiding explicit yarn geometry~\cite{zhu2024realistic,khattar2026texture}, but represent static states rather than deformation-driven changes in spacing and occlusion. More recently, Kalinov \etal~\cite{DeformableTexturesKalinov26} demonstrate deformation-dependent texture content using physics-inspired procedural rules. Their approach prescribes classes of texture behavior under deformation rather than deriving appearance from an underlying deforming microgeometry.

We introduce \textit{\methodname} (\methodabbr), a representation that compiles deformation-driven microgeometry into a continuous, compact neural texture. Our key insight is that when fine-scale geometry is determined by local deformation, its rendering-relevant properties can be represented directly in material space as functions of that deformation. This turns dynamic microgeometry rendering into deformation-conditioned texture evaluation.

We demonstrate \methodabbr on yarn-based textiles. We sample a tileable yarn model across in-plane stretch and shear, map the resulting configurations to a common material-space domain, and render intrinsic maps describing coverage, surface normal, depth, and material identity. A coordinate-based neural model compresses these samples into a continuous representation that is evaluated directly on coarse deforming meshes at runtime. Because its content is learned from the underlying deformed geometry, \methodabbr captures changes in spacing, orientation and visibility, including yarn structures that become exposed or occluded during deformation.

Importantly, \methodabbr represents intrinsic geometric information rather than final radiance. 
Lighting and material response are applied only after evaluating the neural texture, using conventional shading. This separation keeps appearance authorable: lighting, BRDF parameters, and the visual properties assigned to different yarns can be changed without retraining the neural representation.  The texture-space representation also enables neural evaluation to
be decoupled from frame rendering. We cache predicted maps persistently in
material space and selectively update them as deformation changes, allowing
the remaining rendering pipeline to reuse conventional textures across frames
and rendering operations.

In summary, our contributions are: 

\begin{itemize}[noitemsep]
\item \textbf{Deformable Texture Maps}, a continuous, compact neural representation of deformation-driven microgeometry as intrinsic texture-space fields conditioned on local deformation.
\item A \textbf{canonicalization and data-generation pipeline for yarn-based textiles} that samples deformation-dependent yarn geometry, maps each configuration to a shared material-space domain, and renders tileable intrinsic maps for learning.
\item A persistent \textbf{texture-space rendering} framework that decouples neural
evaluation from frame rendering through selective deformation-dependent
updates and cross-frame reuse, while exposing conventional maps to downstream
rendering.

\end{itemize}

We will publish the dataset and models upon publication.

\section{Related Work}
\label{sec:related_work}

\begin{figure*}[!t]
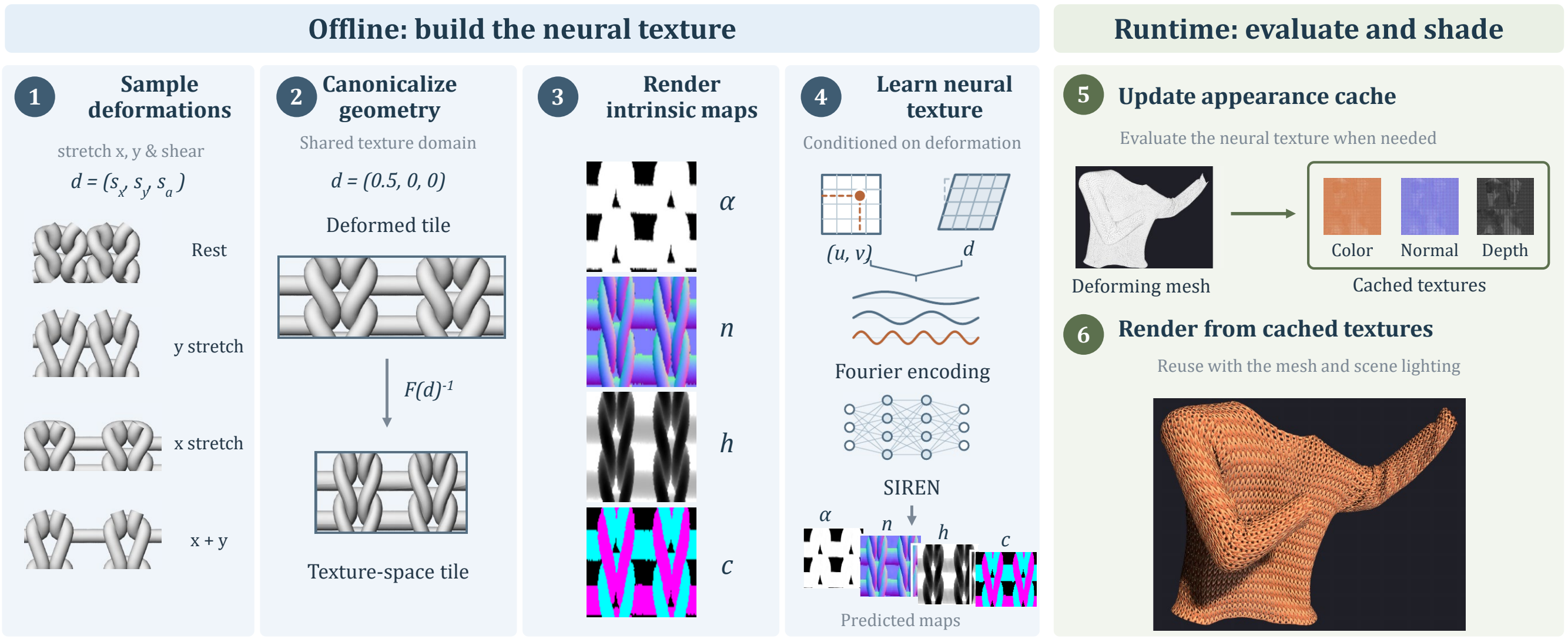

\captionsetup{skip=9.3pt,belowskip=4.55pt}
\vspace*{11pt}
  \centering
\genfig[width=\textwidth]{method/artistic_pipeline_composition_2.pdf} \caption{ 
Overview of Deformable Texture Maps.
Steps 1--3 simulate and render intrinsic maps from detailed microgeometry (\secref{method_data}); Step 4 compiles these maps into a deformation-driven neural texture (\secref{method_model}); and Steps 5--6 evaluate, cache, and shade the predicted maps on a coarse deforming surface at runtime (\secref{method_rendering}).
}
  \label{fig:pipeline_overview}
\end{figure*}
\subsection{Deformation-dependent surface detail.}

A broad class of methods enriches coarse deforming surfaces with fine-scale geometric detail. For skin, wrinkles and other fine-scale features are commonly represented as displacement over a coarser facial surface and driven by facial deformation, strain, or expression \citep{larboulette2004real,ma2008facial,nagano2015skin,feng2021learning,raman2023mesh,chai2023hiface,weiss2024fast}. Related approaches synthesize folds and wrinkles over coarse cloth and garment simulations using procedural, example-based, or learned models \citep{wang2010example,kavan2011physics,zurdo2012animating,muller2010wrinkle}. These methods primarily refine the geometry of the underlying deforming surface.

Our work is inspired by the same coarse-to-fine principle, but targets appearance-driving microgeometry whose structure is distinct from the coarse surface itself. This distinction is particularly important for textiles, where yarns rearrange under deformation and can separate, tighten, or occlude one another. Mechanics-aware methods model these effects explicitly: Montazeri \etal \citep{montazeri2019mechanics} use yarn-level mechanics to drive fiber-level deformation, while Sperl \etal \citep{sperl2021madypg} interpolate precomputed yarn configurations from coarse cloth deformation. Both ultimately recover explicit fine-scale geometry for rendering, whereas we learn a continuous mapping from material coordinates and deformation directly to rendering-relevant fields.

Recent work by Kalinov \etal \citep{DeformableTexturesKalinov26} models deformation-dependent detail directly in texture space using physics-inspired procedural rules, complementary to our approach of deriving texture content from underlying deforming microgeometry.

Beyond skin, deformation-dependent microgeometry has also been studied for textiles. Wang~\etal \cite{wang2010example} and Montazeri~\etal \citep{montazeri2019mechanics} model mechanics-dependent changes in textile microstructure, using yarn-level simulation to drive the deformation of fiber-level geometry. Their work demonstrates how changes in yarn and fiber arrangement under external forces can produce significant changes in macroscopic cloth appearance. 
While their method explicitly generates the resulting fine-scale geometry for rendering, our approach compiles deformation-driven microgeometry into texture maps, avoiding explicit fine geometry at runtime.

Deformation-dependent appearance has also been addressed directly in texture space. Kalinov~\etal\citep{DeformableTexturesKalinov26} introduce physics-inspired procedural textures that respond to surface deformation. In contrast, we derive deformation-dependent texture content from underlying microgeometry, capturing changes in its rendering-relevant properties.

\subsection{Microgeometry-Based Textile Modeling and Rendering}

Explicit yarn-level models capture the mechanics and structure of knitted and
woven fabrics \citep{kaldor2008knit,yuksel2012stitchmeshes,cirio2014yarn,
leaf2018interactivedesign}, but their cost grows with geometric complexity.
Homogenized approaches reduce the simulation cost by recovering yarn-scale
detail from a coarse cloth representation. Homogenized Yarn-Level Cloth
\citep{sperl2020hylc} incorporates yarn-level mechanical response into a
continuum model, while MADYPG \citep{sperl2021madypg} reconstructs detailed
yarn configurations from precomputed deformation-dependent displacement
tables. Our data-generation stage builds directly on MADYPG
(\secref{method_data}). These approaches efficiently recover
deformation-dependent microgeometry, but the resulting yarn geometry must
still be instantiated and rendered explicitly.

Rendering cost has been addressed through complementary representations. Wu and Yuksel \citep{Wu:2017:rtcr} render
fiber-level detail in real time using an efficient yarn-based representation,
while other methods develop physically based appearance models at the fiber,
ply, or yarn scale
\citep{AliagaEGSR2017,Montazeri:2020:PlyBasedCloth}.
More recent surface-based models avoid explicit textile microgeometry
altogether by reproducing fine-scale woven or knitted appearance directly on a coarse
surface \citep{zhu2024realistic,khattar2026texture}. 

Recent work further demonstrates that deformation-dependent yarn geometry can be reconstructed and rendered at interactive rates \citep{huang2025real}. Such approaches retain explicit 3D microgeometry at runtime, preserving its geometric expressiveness while keeping reconstruction and rendering coupled to the represented geometry. DTM explores a complementary tradeoff: when local microgeometry is determined by coarse deformation, we compile its rendering-relevant properties into a deformation-conditioned texture representation. This replaces runtime reconstruction and processing of explicit yarn geometry with texture evaluation on the coarse surface, while allowing the resulting maps to integrate with conventional texture-based rendering pipelines.

\subsection{Neural material and appearance representations.}

Neural networks have been used to compress and interpolate measured reflectance \citep{rainer2019btf},
represent multi-resolution material appearance \citep{kuznetsov2021neumip},
and compress PBR texture sets into compact neural
representations that can be decoded efficiently at runtime
\citep{weinreich2024real,laurent2025hardwareacceleratedneuralblock}.
Other approaches learn neural models of cloth appearance \citep{soh2024neuralappearancemodelcloth,chen2024rtwoven},
or serve as real-time appearance models inside production renderers \citep{zeltner2024neural}.
These methods focus on static material appearance or texture
content. In contrast, DTM represents how fine-scale structure changes with
deformation: the neural texture is conditioned on local deformation and
predicts texture maps derived from the corresponding microgeometry.

The term \emph{neural texture} comes from deferred neural rendering \citep{thies2019deferred}, which stores a learned feature map on a surface and decodes it with a convolutional renderer. We keep the term for a learned, UV-indexed texture defined by a neural network, but condition it on deformation and directly output texture maps rather than learned features, allowing the result to be consumed by conventional rendering pipelines.

\section{\methodname}
\label{sec:method}

\paragraph*{Overview}
\label{sec:method_overview}

Our method replaces runtime reconstruction and rendering of deformation-dependent microgeometry with neural evaluation of deformation-conditioned intrinsic neural fields. Assuming that fine-scale geometry is determined by local deformation, we generate detailed geometry offline, map it to a common material-space domain, and compile its rendering-relevant properties into a continuous neural texture \(\neuraltexture\). We formalize \(\neuraltexture\) as
\begin{equation}
  \neuraltexture : [0,1]^2 \times \mathcal{D} \to \R^{8},
  \qquad
  (\vv{u}, \vv{d}) \mapsto (\alpha, \vv{n}, h, \vv{c}),
  \label{eq:neural_texture}
\end{equation}
where \(\mathbf u\in[0,1]^2\) denotes a coordinate in the canonical material domain and \(\mathbf d\in\mathcal D\) denotes the local deformation state ($\mathcal D \simeq \mathbb{R}^3$, see Sec.~\ref{sec:method_data}). The outputs comprise opacity \(\alpha\) -- derived from the fractional coverage of the microgeometry, surface normal \(\mathbf n\), depth \(h\), and a three-channel material-identity code \(\mathbf c\). Because these fields encode geometry rather than radiance, lighting and material response remain separate from the learned representation and can be modified without retraining.

An overview of the method is presented in Figure~\ref{fig:pipeline_overview}. We instantiate this representation for yarn-based textiles using the homogenized yarn model of Sperl~\etal \citep{sperl2021madypg}. First, we sweep a yarn patch through deformation space, reconstruct and canonicalize the resulting microgeometry, and render the target intrinsic fields (\secref{method_data}). Second, we fit the deformation-conditioned neural texture \(\neuraltexture\) to these samples (\secref{method_model}). Finally, at runtime, we evaluate \(\neuraltexture\) on a coarse deforming cloth mesh and shade the predicted fields using conventional material and lighting models (\secref{method_rendering}).

\begin{figure}[!t]
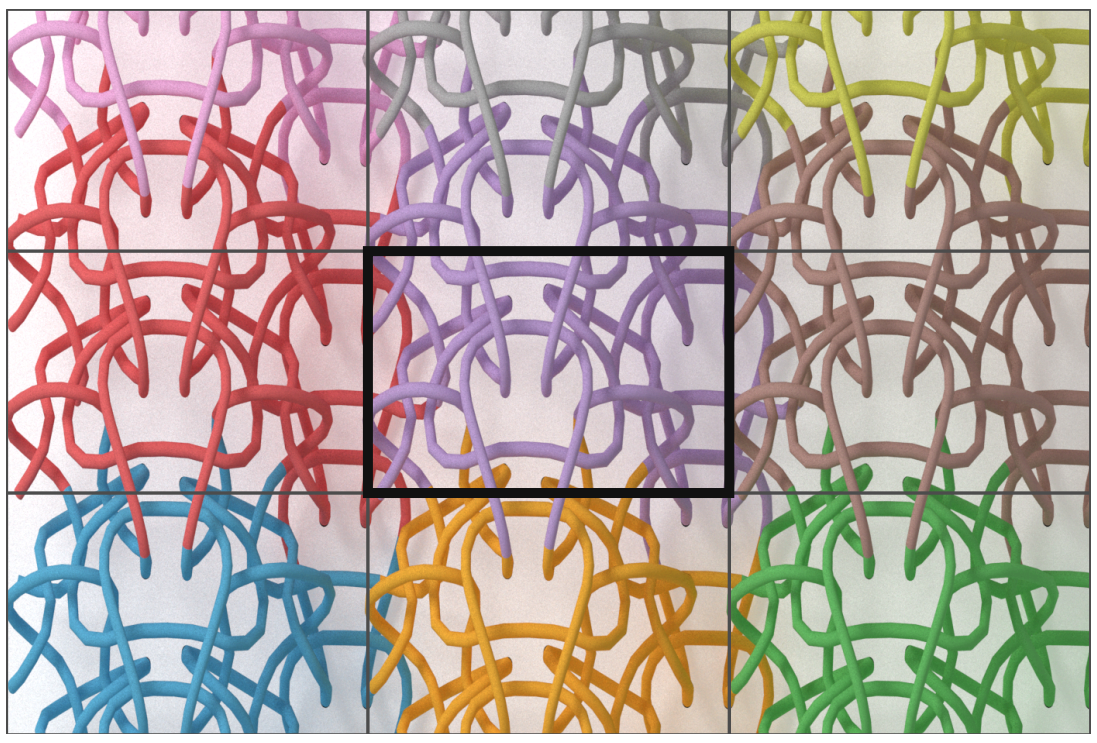

\captionsetup{skip=9.3pt,belowskip=9.3pt}
  \centering
  \genfig[width=0.85\columnwidth]{method/yarn_tile_overlap_rest.pdf}
  \caption{Rest-pose yarn splines replicated over a $3\times3$ tile neighborhood.
  The black outline marks the central tile retained after rendering and cropping;
  the surrounding tiles preserve yarn geometry that crosses the central tile's
  boundaries. %
  }
  \label{fig:yarn_tile_overlap_rest}
\end{figure}

\subsection{Microgeometry Compilation}
\label{sec:method_data}

We generate data for \(\neuraltexture\) by sweeping a tileable yarn configuration through deformation space and converting each resulting microgeometry state into a common material-space representation. For each deformation state, we simulate the yarn configuration, construct renderable geometry, canonicalize it to the rest-space tile footprint, and render the resulting intrinsic fields.

\paragraph*{Simulation}

Starting from a planar tile containing the rest-pose yarn splines, we sample a Cartesian grid of deformation states $\vv d=(s_x,s_y,s_a) \in [m,M]^2\times[-1,1]$ ($m/M$ being the minimum/maximum stretch values captured during simulation). Each state applies the in-plane deformation gradient
\begin{equation}
  \mat{F}(\vv d) =
  \begin{pmatrix}
    1 + s_x & s_a \, (1 + s_y) \\
    0       & (1 + s_y) \sqrt{1 - s_a^2}
  \end{pmatrix},
  \label{eq:deformation_gradient}
\end{equation}
Defining \(s_a\) as the %
sine of the shear angle (\(s_a = sin(\alpha)\)) keeps the norm of the second column equal to \(1 + s_y\), allowing shear to change its orientation without introducing additional stretch. %
We use $\mat{F}(\vv d)$ to prescribe the macroscopic deformation and obtain the corresponding relaxed yarn configuration with MADYPG \citep{sperl2021madypg}, running physical relaxation until equilibrium. Each resulting configuration therefore associates a deformation state $\vv d$ with a detailed set of yarn centerlines.

\paragraph*{Meshing}
MADYPG yields periodic deformed yarn centerlines as a result of homogenization. To obtain renderable surface geometry, we construct a generalized cylinder with a fixed circular cross-section around each rest-pose centerline. For each deformation state, we displace its vertices by the rest-to-deformed centerline offsets, 
 expressed in the 
adapted frame featuring the smallest torque along the centerline. We enforce consistency of this frame across the tile boundaries so that the resulting surface geometry remains periodic and tileable.
To preserve geometry crossing the periodic boundaries during rendering, we replicate the resulting yarn tile over a \(3\times3\) neighborhood, render it orthographically, and retain the central tile (Figure~\ref{fig:yarn_tile_overlap_rest}). The surrounding copies provide the portions of yarn geometry crossing into the central tile, ensuring seamless boundaries.

\begin{figure}[!t]
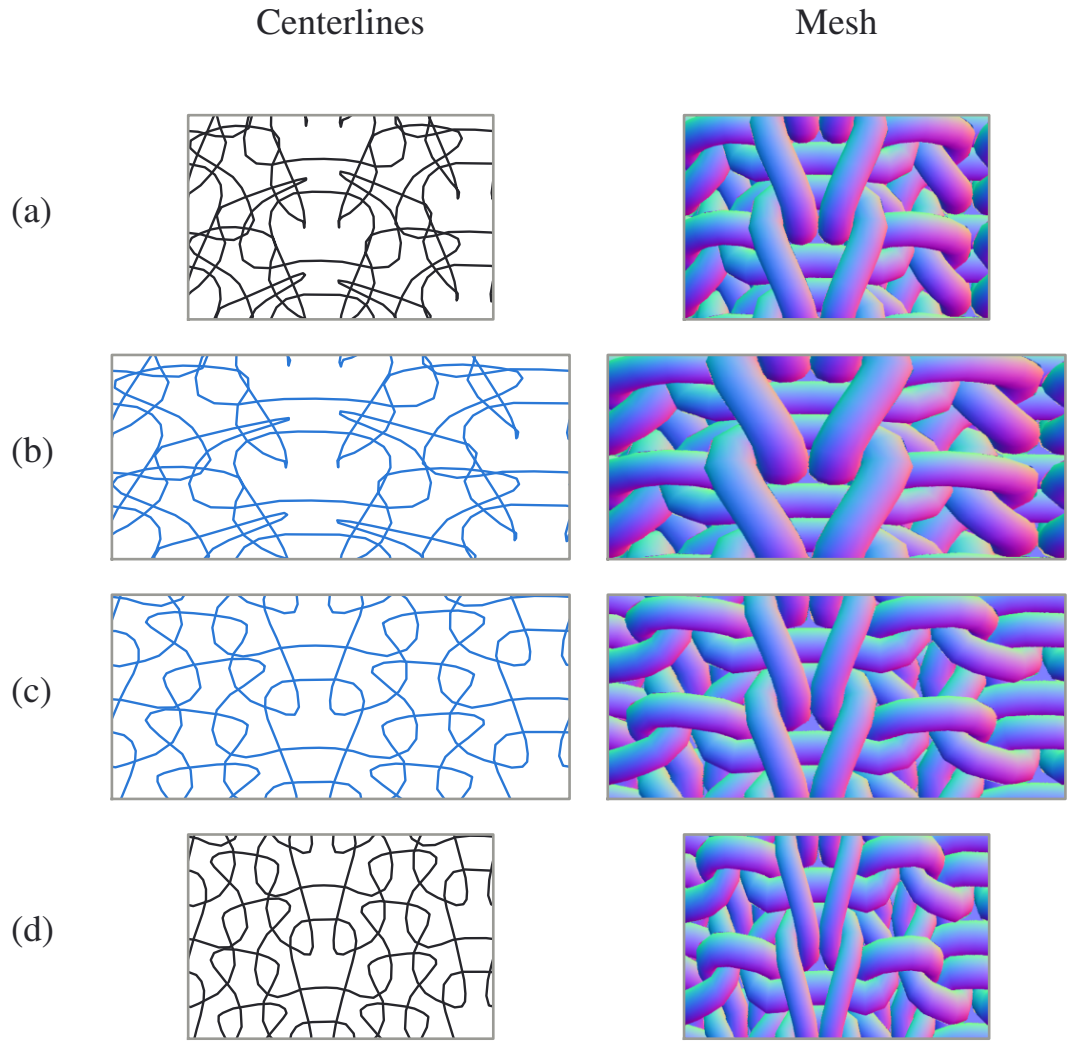

\captionsetup{skip=9.3pt,belowskip=9.3pt}
  \centering
  \genfig[width=0.85\columnwidth]{method/pullback_process_mock.pdf}
  \caption{Illustration of the canonicalization process. (a) Rest pose. (b) The rest geometry after linearly stretching it to the target deformation state, without yarn-level simulation. (c) The simulated yarn deformation in world space. (d) The simulated geometry pulled back into material space. In each row, the left column depicts yarn centerlines and the right column the corresponding meshes.%
  }
  \label{fig:pullback_correction}
\end{figure}

\paragraph*{Canonicalization}
The deformed yarn geometry has a spatial footprint that varies with the deformation state \(\mathbf d\), whereas our neural texture is defined over a fixed material-space domain. We therefore pull each configuration back to the rest-space tile by applying \(\mat F(\mathbf d)^{-1}\) to the deformed geometry before rasterization. This transformation canonicalizes the sampling domain without undoing the simulated yarn rearrangement, preserving deformation-dependent changes in spacing, occlusion, and visibility. As a result, all deformation states are represented over a common texture-space footprint and sampled at a consistent spatial density. This is illustrated in Figure~\ref{fig:pullback_correction}.

\paragraph*{Rendering}

We rasterize each canonicalized yarn configuration into the intrinsic maps used to train \(\neuraltexture\).The render pass produces four maps indexed by the canonical texture coordinate \(\mathbf u\): alpha \(\alpha\), representing fractional yarn opacity coverage $\alpha$; surface normal \(\mathbf n\); depth \(h\); and material identity \(\mathbf c\).

Although positions are rasterized in the canonical domain, the normal map should describe the yarn surface in the deformed configuration. We therefore lift the in-plane deformation to three dimensions as \(\widetilde{\mat F}=\operatorname{diag}(\mat F,1)\), apply \(\widetilde{\mat F}\) to the canonicalized mesh vertices, and recompute smooth vertex normals on the resulting deformed mesh. These normals are interpolated at the corresponding rasterized surface points and stored as the target normal map. Thus, canonicalization provides a common spatial domain while retaining the deformation-dependent surface orientation used for shading.

Finally, each texel together with its deformation state defines a training sample \((\vv{u}, \vv{d}, \alpha, \vv{n}, h, \vv{c})\).

\subsection{Deformation-Conditioned Neural Texture}
\label{sec:method_model}

Recent neural texture compression methods combine learned latent textures with lightweight MLP decoders to reconstruct material maps efficiently at runtime \citep{laurent2025hardwareacceleratedneuralblock,laurent2026neuraltexturecompressionusing}. In our setting, however, the target maps must additionally vary continuously with deformation. We therefore directly parameterize the joint material-coordinate and deformation domain with a coordinate-based network
\citep{sitzmann2020siren,tancik2020fourier}. To model high-frequencies we combine a fixed Fourier encoding with sinusoidal activations. We compare this choice against alternative coordinate-network \cite{gao2024hsirenimprovingimplicitneural} \cite{alsakabi2026fmsirenfmfinerimplicit} architectures in supplemental material. %
Let $ \vv p = ( \vv u,\vv d )$ denote the concatenated texture coordinate and deformation state. Our neural texture is
\begin{equation}
  \neuraltexture = S \circ \gamma,
  \qquad
  (\alpha, \vv{n}, h, \vv{c}) = S\bigl(\gamma(\vv p)\bigr),
  \label{eq:model}
\end{equation}
where  $\gamma$ independently maps each component \(\vv{p}\) to
\begin{equation}
  \gamma(\vv{p}) = \bigl(\vv{p},\; \sin(2^{\ell} \pi \vv{p}),\; \cos(2^{\ell} \pi \vv{p})\bigr)_{\ell = 0}^{L-1}.
  \label{eq:encoding}
\end{equation}
The resulting features are concatenated and processed by the SIREN \(S\), which jointly predicts the eight channels of the four target maps. %

Fourier features expose multiple frequency bands directly to the network, complementing the periodic activations of SIREN. We evaluate the contribution of this encoding against a bare SIREN at matched parameter budgets in the supplemental material.

We train all maps jointly using
\begin{equation}
  \mathcal{L} = \sum_{f \in \{\alpha, \vv{n}, h, \vv{c}\}} \mathrm{MSE}\bigl(\hat{f}, f\bigr),
  \label{eq:loss}
\end{equation}
Here, \(\hat{f}\) denotes the network prediction of target map \(f\) at the same texture
coordinate and deformation state. Each MSE is averaged over the channels of its map, giving equal weight to the four maps losses. Material identity is represented by a continuous three-dimensional code \(\vv c\); at inference, predictions are decoded to the material ID associated with the nearest codeword.

\begin{figure*}[!t]
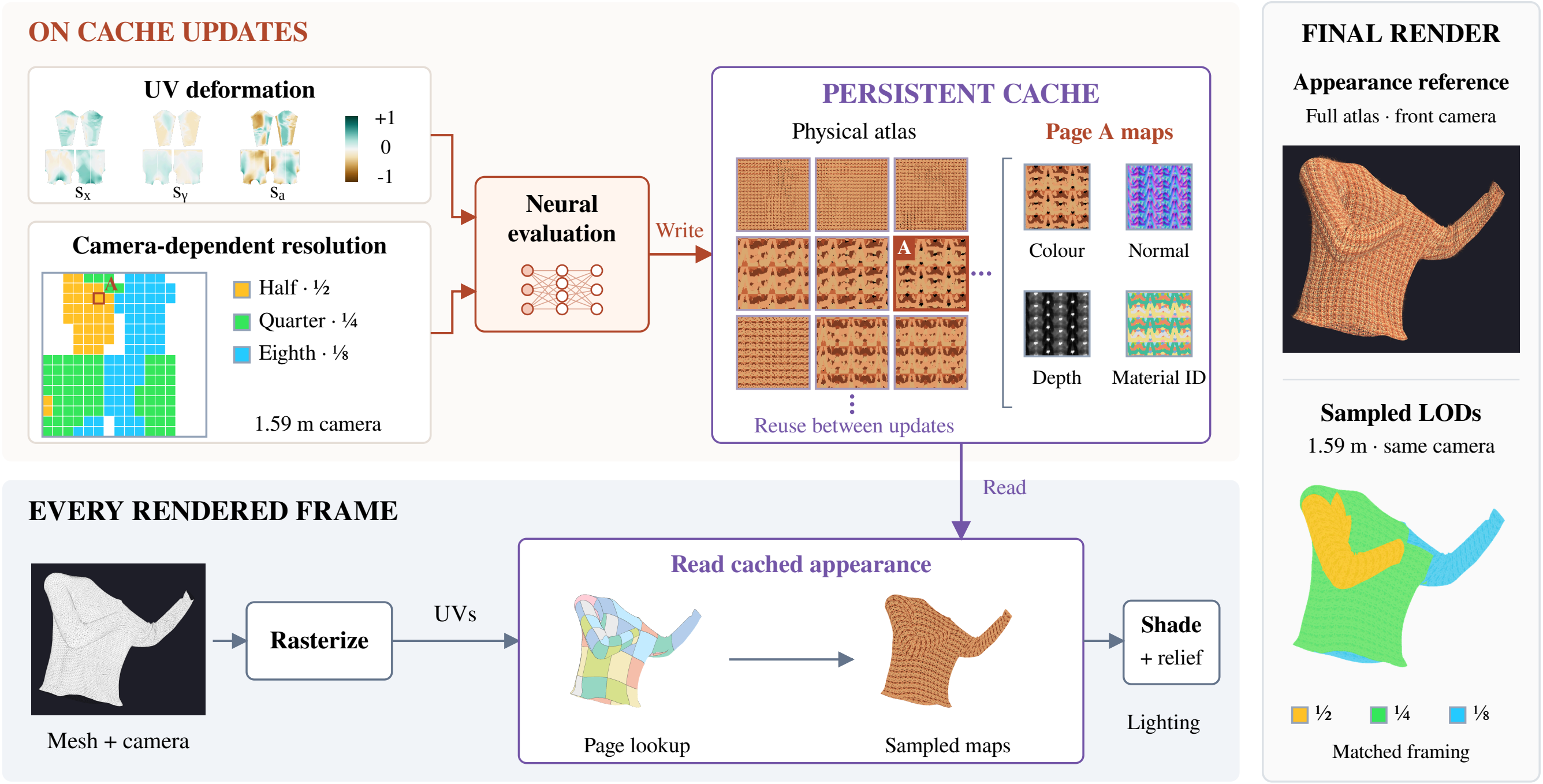

\captionsetup{skip=9.3pt,belowskip=4.55pt}
\vspace*{11pt}
\centering
\genfig[width=\textwidth]{method/texture_cache_rendering.pdf}
\caption{Persistent texture-cache rendering.
UV deformation $(s_x,s_y,s_a)$ and view-dependent detail requests drive neural
updates; each frame reuses cached maps for shading and relie.
Page A shows the stored maps (depth normalized, material IDs colour-coded).
Right: a full-atlas reference render and a coarse-mesh LOD diagnostic from the
same camera.}
\label{fig:texture_cache_rendering}
\end{figure*}

\section{Runtime Rendering}
\label{sec:method_rendering}

The simplest integration of DTM evaluates the neural texture on demand in the fragment shader.
While straightforward, our early experiments showed that this unoptimized strategy results in poor performance.
A more advanced strategy is to query the network only at visible surface samples every frame, using deferred shading. The coarse mesh is rasterized into screen-space buffers (a G-Buffer) containing texture coordinates, per-vertex deformation parameters $(s_x,s_y,s_a)$, and a coverage mask. 
The network is then evaluated in GPU batches over covered pixels, and the predicted opacity, normals, depth, and material ID are stored for shading in a subsequent pass.
This substantially reduces redundant neural evaluations, but remains coupled to screen-space visibility and inherits limitations of deferred shading for effects requiring multiple or partially occluded surface samples, such as transparency.
We instead adopt a persistent texture-space strategy inspired by decoupled shading~\cite{baker2022generalized}, separating neural evaluation from frame rendering, and sparse virtual textures~\cite{barrett2008sparsevirtualtextures}

\subsection{Texture-Space Caching}
\label{render:integration}
Our primary rendering strategy follows \cite{baker2022generalized} and exploits the UV parameterization of the DTM to cache network predictions in texture space (Figure~\ref{fig:texture_cache_rendering}). As the predicted material maps are independent of lighting and viewing direction, they can be reused across frames and selectively updated as the cloth deforms. This trades texture memory for fewer neural evaluations and decouples network evaluation from per-frame rendering.

To obtain deformation inputs at texture resolution, we rasterize the coarse mesh in UV space, interpolating the per-vertex parameters $(s_x,s_y,s_a)$ into a deformation texture. For each texel selected for neural evaluation, we sample this texture with bilinear filtering and combine the resulting deformation state with the local yarn-texture coordinate as input to the network. The resulting predictions are stored directly in material textures.

We decompose the UV space into pages, to which we assign a constant Level of Detail (LoD, see "Camera-dependent resolution" in Figure~\ref{fig:texture_cache_rendering}), which allows us baking high-resolution texture outputs in close-up regions while remaining memory friendly.
To ensure an adequate final texel ratio on screen after shading, we determine a target resolution for each triangle based on its projected area and the screen resolution. All triangles are processed in a first pass, and each UV-page is assigned with a rasterization LoD corresponding to the maximum of selected LoDs for all triangles projecting on said page. Triangles falling outside the camera frustrum are marked as discarded: they do not influence the pages' LoDs and are not baked, as they are not visible on screen. Once the pages' LoDs are set up, the neural texture is baked at adapted resolution over the UV-space.

Representing the network output as dynamically updated textures allows standard rendering techniques to access the predicted maps without further neural evaluations. In particular, we use forward rendering to handle transparency, while parallax mapping, flyaway fibers, and fuzz shells repeatedly reuse the cached maps. Table~\ref{tab:rendering_path_capabilities} details the differences between the two rendering strategies.

\begin{table}[!htbp]
\centering
\caption{Rendering capabilities of the two implemented paths.}
\label{tab:rendering_path_capabilities}
\small
\setlength{\tabcolsep}{3pt}
\renewcommand{\arraystretch}{1.15}
\begin{tabularx}{\columnwidth}{@{}>{\raggedright\arraybackslash}X
  >{\raggedright\arraybackslash}X >{\raggedright\arraybackslash}X@{}}
\toprule
\textbf{Capability} & \textbf{Our Approach} & \textbf{Screen-space deferred} \\
\midrule
Cross-frame reuse & Persistent UV entries & Per-frame reevaluation \\
\addlinespace[2pt]
Local deformation updates & Refresh affected regions & Evaluate covered pixels \\
\addlinespace[2pt]
Independent update rate & Reuse between updates & Coupled to frame evaluation \\
\addlinespace[2pt]
Material-space effects & Repeated texture lookups & Missing samples require evaluation \\
\addlinespace[2pt]
Transparent layers & Texture lookup per layer & Predictions needed per layer \\
\bottomrule
\end{tabularx}
\end{table}

\subsection{Cache Update Strategy}
\label{render:cache}

Because the cache stores texture maps, changes in lighting or viewpoint do not by
themselves require new neural predictions at a fixed texture resolution. We exploit this
reuse with a deformation-triggered update policy and a configurable update schedule.

\paragraph*{Deformation-triggered invalidation.}
For each coarse-mesh vertex \(v\), we compare its current deformation
\(\vv d_v=(s_x,s_y,s_a)\) with a stored reference \(\widehat{\vv d}_v\).
We mark the affected texture regions for reevaluation when
\[
  \lVert \vv d_v-\widehat{\vv d}_v\rVert_\infty > \tau,
\]
where \(\tau\) is a tolerance in the original deformation coordinates, without
per-component normalization. On a threshold crossing, we advance the vertex's reference
to its current deformation and mark all row groups in the affected regions as pending.
Selecting one row group for an update leaves the others pending; the reference therefore
records a detected change, rather than a common state of all cached texels. Comparing
against this reference allows small successive changes to trigger an update once their
net departure exceeds \(\tau\).

\paragraph*{Update scheduling.}
We set a target update frequency \(f_u\) independently of the rendering frame rate and
cycle through \(G\) groups of texture rows during motion. With \(G=2\), successive
scheduled updates alternate between even and odd rows. If updates occur at the target
rate without interruption, each group receives an update opportunity every \(G/f_u\)
seconds. This is a nominal scheduling interval, not a bound on prediction age: thresholding
can suppress updates, and batching or deferred submissions can delay them. Initialization
and forced refreshes bypass row interleaving; a settling refresh also uses full row
coverage before unchanged deformation can be reused without further updates. The
supplementary material details these exceptions and the runtime settings.

\paragraph*{Quality and responsiveness.}
Skipped texels retain their cached intrinsic fields, while their rendered appearance can
still change with lighting, viewpoint, and coarse-mesh motion. The tolerance \(\tau\)
controls deformation-space differences and does not bound appearance error. Reduced update
frequency and row interleaving can delay changes in the predicted yarn fields and leave
neighboring rows representing different deformation states. Evaluating these approximations
requires comparison with full per-frame texture-space updates at the same resolution, so
that errors due to update scheduling are separated from representation and sampling errors.

\subsection{Yarn Shading and Cloth Effects}
\label{render:shading}

The cached neural maps are used as conventional material textures to render the coarse cloth. Alpha determines yarn coverage, normals provide local surface orientation, and the material ID selects the corresponding shader parameters. We use Lambertian diffuse and GGX specular reflection, while depth and alpha support view-dependent relief and local occlusion.

Our maps-based representation enable additional yarn-scale effects without further neural evaluations. Parallax mapping uses depth to produce view-dependent yarn relief, while fuzz is approximated with thin shell layers anchored to the yarn surface. Sparse flyaway fibers are represented as tapered ribbons, with alpha selecting their roots, depth positioning them on the yarns, and normals guiding their orientation \cite{Lengyel:2001:RFO}. Further implementation details are provided in the supplementary material.

\resulttable{figure7_grouped_fields}{
Reconstruction fidelity across five yarn structures using a width of 128, we measure PSNR (dB) for alpha, normals and depth and report the proportion of material IDs that are correctly reconstructed. For all columns, higher is better ($\uparrow$).
}

\resulttable{storage_quality_fields}{
Storage cost and reconstruction quality
on \textsc{rib}. We compare our neural models with ASTC block compression
of the ground-truth textures and lossless texture references, illustrating
the trade-off between storage and representation quality PSNR of in-distribution samples. We compare different widths of our Fourier-SIREN network, among which best scores are bold, second best underlined; we highlight the trade-off that we use in practice.
}

\section{Results}
\label{sec:results}

In this section we evaluate \methodabbr accuracy and performance (Section~\ref{sec:results_accuracy}), appearance control (Sections~\ref{sec:results_garments} and~\ref{sec:results_editability}), and end-to-end rendering performance (Section~\ref{sec:results_runtime}).

\paragraph*{Experimental Setup.} We evaluate \methodabbr on five yarn structures
spanning woven and knitted configurations: \textsc{rib}, \textsc{basket},
\textsc{stockinette}, \textsc{honeycomb}, and \textsc{satin}. Unless stated otherwise,
each structure uses an independently trained Fourier-SIREN of width 128, trained for
30,000 steps on a discrete lattice of deformation states. All models sample an \(8 \times 8 \times 8\) lattice of deformation
states with \(s_x,s_y\in[-0.2,1.0]\) and \(s_a\in[-0.7,0.7]\) for a total of 512 states. Full implementation details
are provided in the supplementary material.
Throughout the quantitative tables, alpha, normals, and depth are measured by PSNR
(dB), and material ID by classification accuracy (\%). Normals are also evaluated qualitatively and quantitatively using lambertian renders and FLIP metric ~\cite{andersson2020flip}.

\begin{figure*}
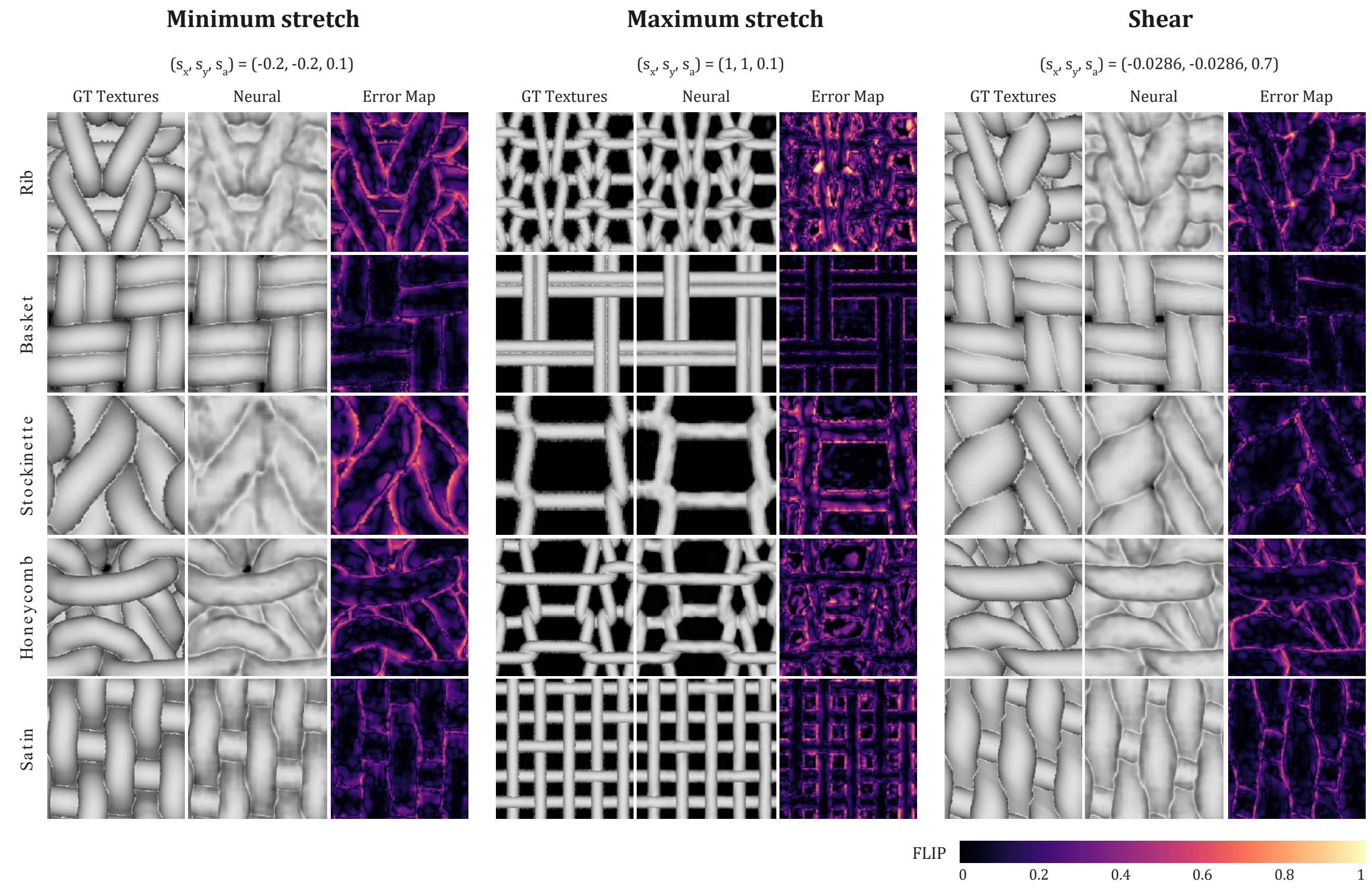

\captionsetup{skip=9.3pt,belowskip=4.55pt}

\centering
\genfig[width=0.99\linewidth]{results/figure7_deformation_composition_2.pdf}
\caption{Reconstruction across five textile structures and three deformation states per structure. %
Columns show a lambertian render using ground-truth intrinsic normals extracted from explicit yarn geometry, a render using neural-predicted normals, and FLIP error between the two map-based renders. Predictions preserve the overall yarn layout, with errors most visible at boundaries and openings.}
\label{fig:figure7_yarn_comparison}
\end{figure*}

\begin{figure}
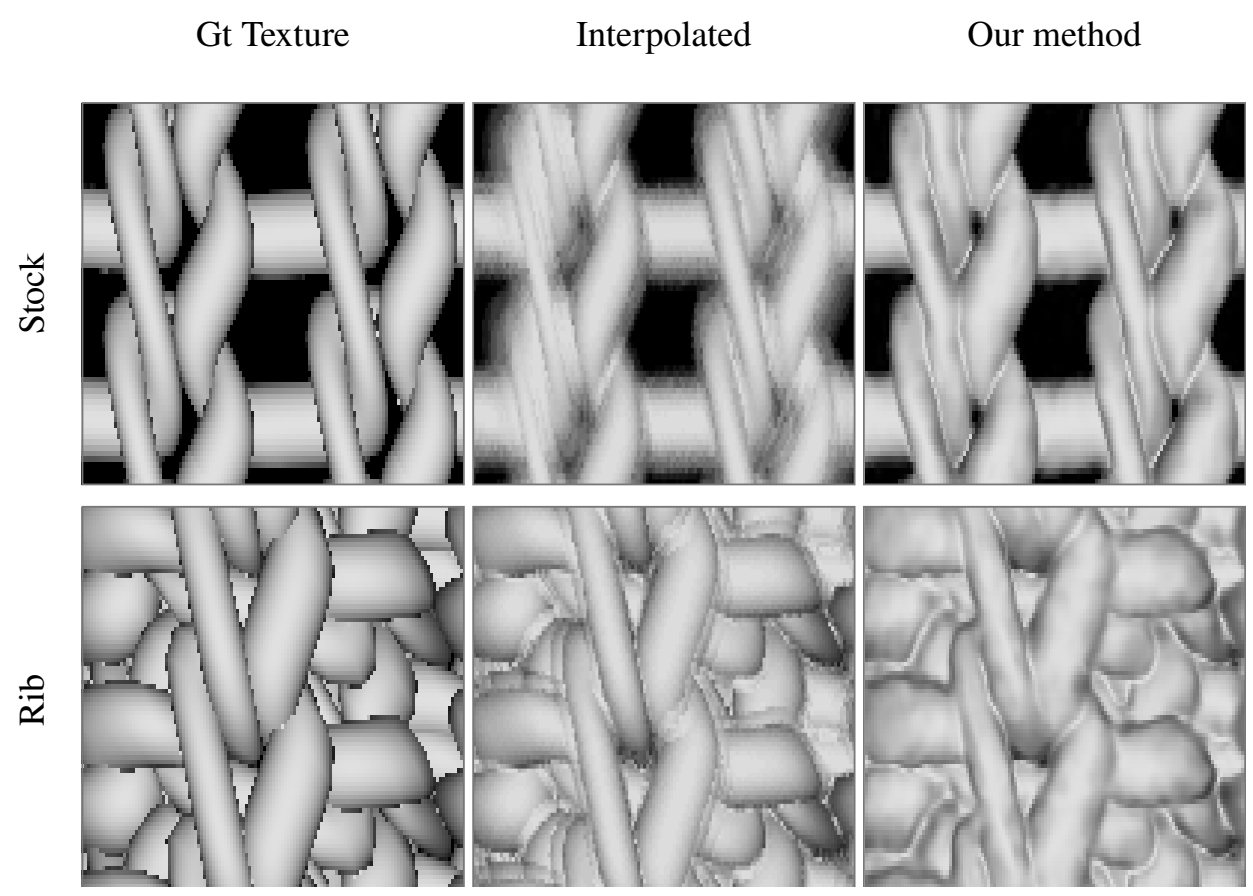

\captionsetup{skip=9.3pt,belowskip=0pt}

\centering
\genfig[width=0.97\linewidth]{results/figure8_aov.pdf} \caption{
Shaded normal comparisons at deformation
states out of distribution for our given dataset. %
Compared with direct geometry rendered @2k (left), interpolated ground-truth textures @128px
(middle) show ghosting at moving yarn boundaries. Our neural predictions @128px
(right) reduce ghosting but smooth fine detail. }
\label{fig:figure8_aov}
\end{figure}

\subsection{Model Accuracy and Performance}
\label{sec:results_accuracy}

We evaluate reconstruction accuracy across five yarn structures and sampled deformation states. 
We compare against the
ground-truth (GT) maps, which are estimated from the original geometry.

Table~\ref{tab:figure7_grouped_fields} reports accuracy for each map. 
Depth and material identity achieve the highest reconstruction accuracy, while alpha and normals 
have less precision. 
Figure~\ref{fig:figure7_yarn_comparison} shows these differences qualitatively by shading the predicted normals with a Lambertian model. The error is measured using FLIP\cite{andersson2020flip}. 
\methodabbr preserves the overall yarn structure even under large deformations. %

We show in Table \ref{tab:storage_quality_fields} how \methodabbr performs at different model sizes. We compare per-map quality against texture compression approaches such as ASTC, performed over the entire training dataset (See supplementary for further evaluations).
Using our selected parameters (Fourier SIREN width set to 128, yellow row), our approach provides quality between the ones of ASTC $(8\times8)$ and ASTC $(12\times12)$, for a 42x-to-94x storage memory reduction.

In Figure~\ref{fig:figure8_aov}, we also compare our method with a na\"ive baseline that blends in image space the neighboring stored textures. 
Linear texture blending superimposes yarn boundaries from neighboring states, producing visible ghosting, whereas \methodabbr produces a more coherent intermediate configuration closer to the ground truth.

\begin{figure}
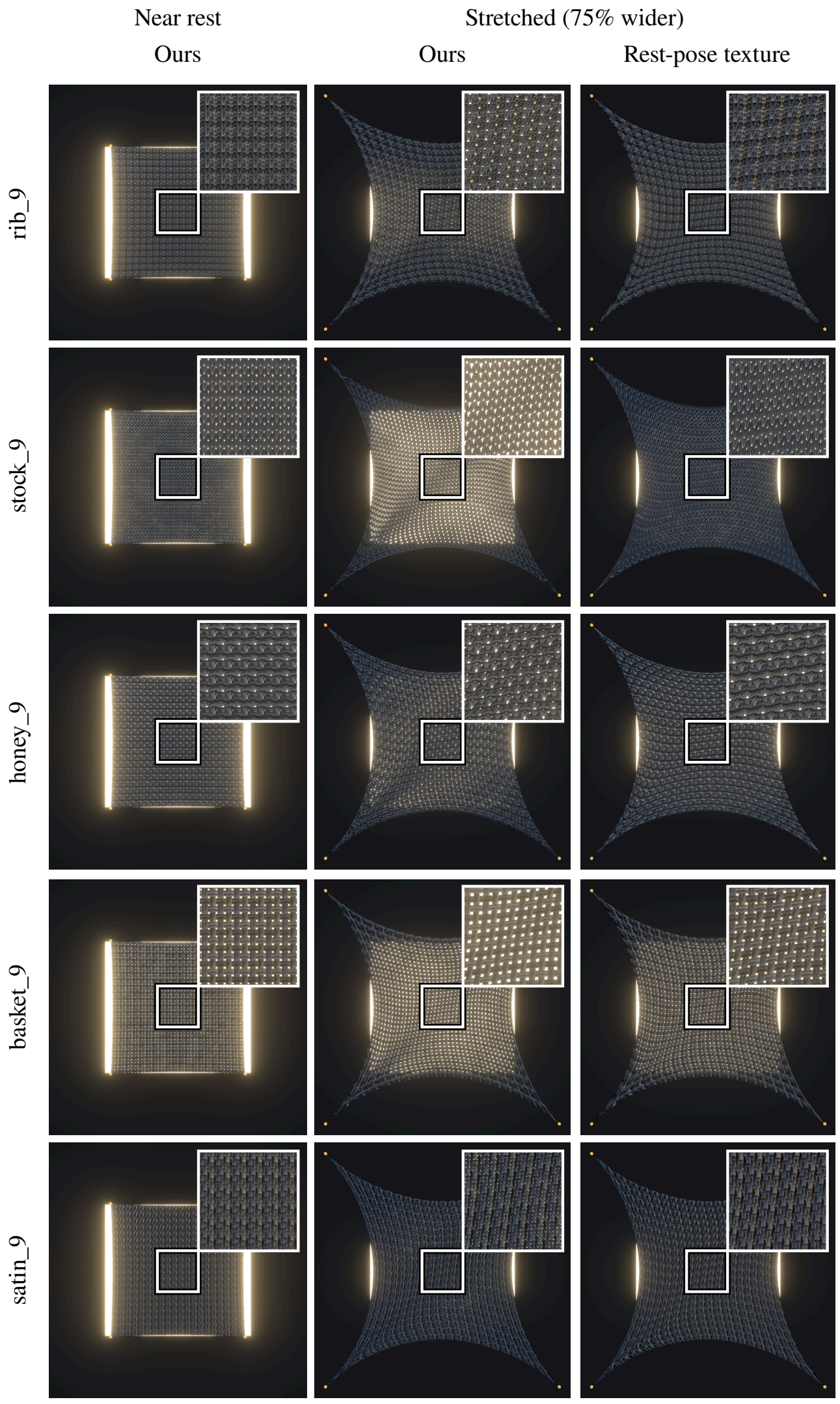

\captionsetup{skip=9.3pt,belowskip=6pt}
\vspace*{11pt}

\centering
\genfig[width=\columnwidth]{results/backlit_cross_yarn.pdf}
\caption{ %
Deformation-dependent yarn coverage for five yarn structures under backlighting. Our method updates coverage
as the cloth stretches, whereas a fixed texture rendered from undeformed yarn geometry
stretches with the cloth UVs. %
}
\label{fig:backlit_cross_yarn}
\end{figure}

\subsection{Garment-level Examples}
\label{sec:results_garments}

We evaluate \methodabbr on simulated coarse garment meshes and cloth patches.
We compute per-triangle stretch and shear in local material coordinates,
average these quantities at mesh vertices using area weights, and interpolate
the resulting deformation parameters across the surface to condition the
tileable neural material.
Figure~\ref{fig:backlit_cross_yarn} illustrates the importance of
deformation-dependent coverage: stretching alters the yarn arrangement and
the openings through which the backlight is visible.
Unlike naive texture scaling, \methodabbr captures these structural changes
through its deformation-conditioned opacity field.
The resulting effect reflects changes in yarn coverage, rather than light
transmission through the yarns.

\subsection{Editability}
\label{sec:results_editability}

\begin{figure}
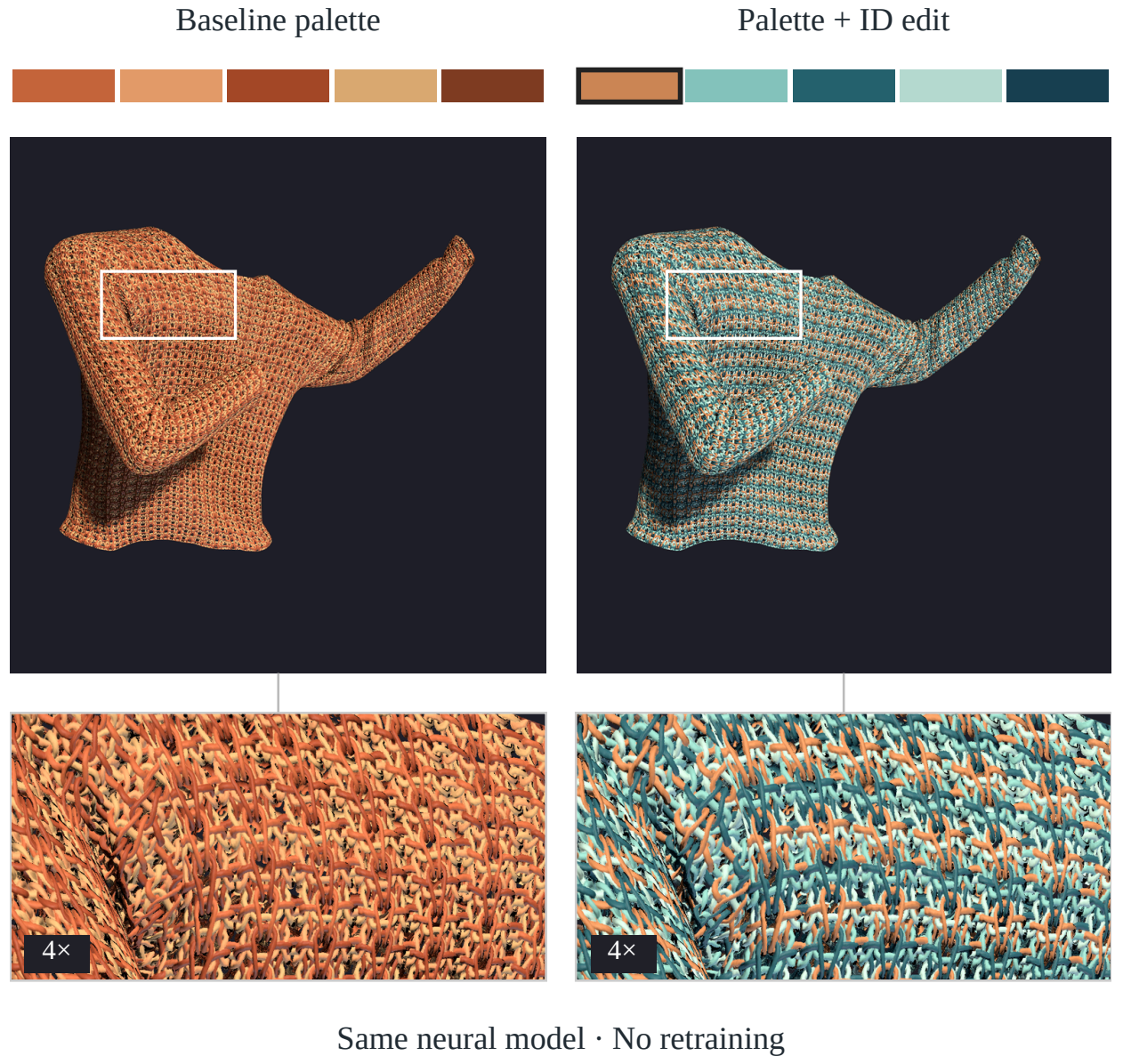

\captionsetup{skip=9.3pt,belowskip=4.42pt}

\centering
\genfig[width=0.97\linewidth]{results/material_id_editing_column.pdf} \caption{Yarn-colour
editing. Left: the baseline palette. Right: a different palette
with one material-ID group edited separately (outlined swatch). The edit follows that
group wherever it appears on the garment. Bottom crops
show matching regions at \(4\times\).}
\label{fig:material_id_editing}
\end{figure}

\begin{figure}
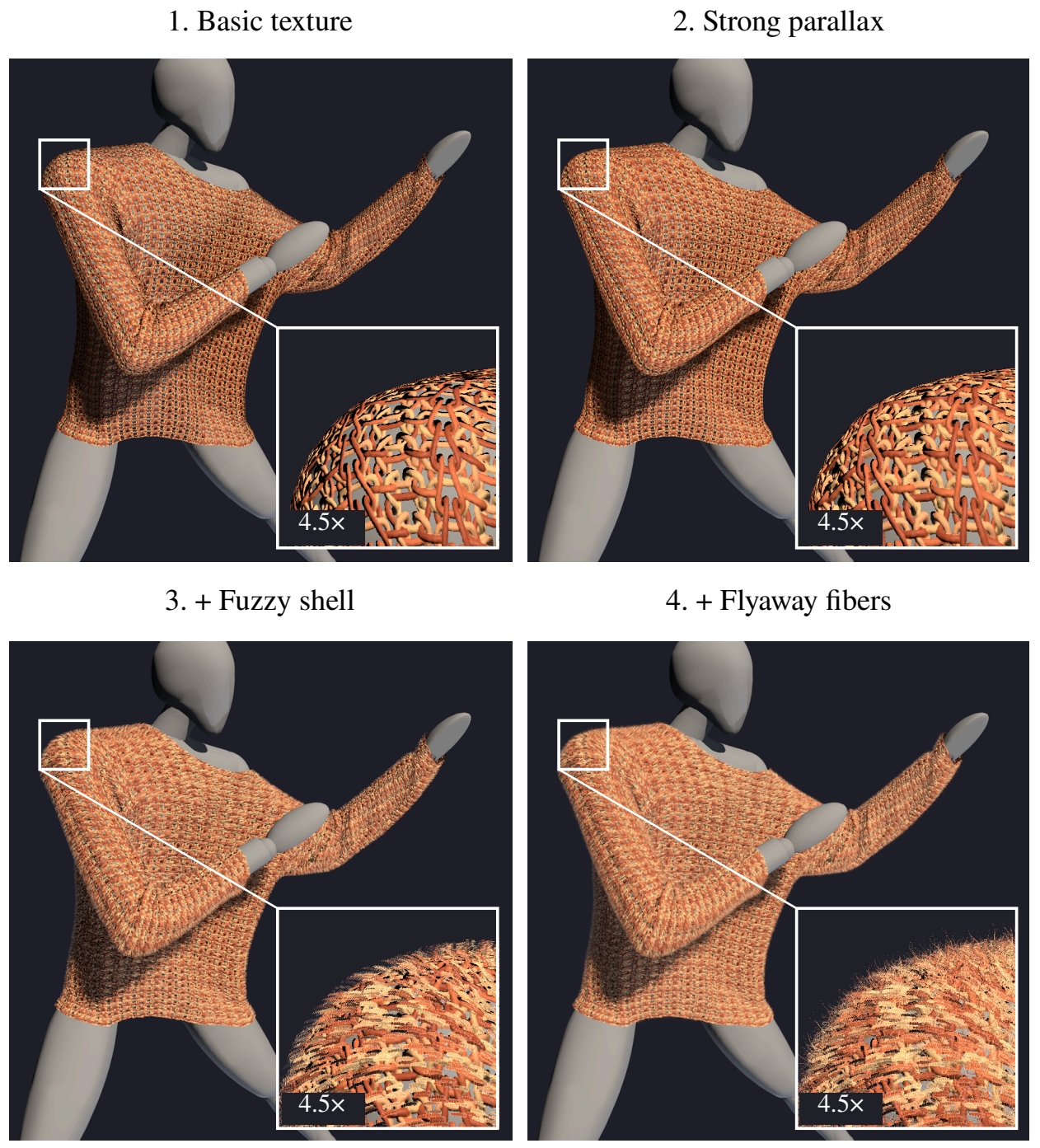

\captionsetup{skip=9.3pt,belowskip=0pt}

\centering
\genfig[width=0.97\linewidth]{results/vfx_parallax_column.pdf} \caption{Effects accumulate
on the same neural garment: (1) basic texture without parallax, (2) added parallax relief,
(3) added fuzzy shell, and (4) added flyaway fibers. Insets show the same shoulder region at \(4.5\times\) magnification,
highlighting the effects along the garment silhouette.}
\label{fig:vfx_buildup}
\end{figure}

Because \methodabbr predicts intrinsic maps rather than final radiance, lighting and material
appearance can be edited without retraining.
Figure~\ref{fig:material_id_editing} shows how predicted material identities enable palette edits 
that consistently follow yarn groups across the textile. The predicted maps can also be 
combined with conventional rendering effects, including parallax relief, fuzzy shells, and flyaway fibers,
 without modifying the learned representation (Figure~\ref{fig:vfx_buildup}). We also demonstrate real-time relightability and pallette edits with fixed models in various animated scenarios (see Teaser \ref{fig:teaser}).

\subsection{Real-Time Renderer Evaluation}
\label{sec:results_runtime}

We evaluate the complete renderer to measure how texture-space reuse affects performance
under deformation and camera motion. We compare per-fragment screen-space deferred neural
evaluation with our texture-space approach, using the same neural
model, shading settings, and \(1920 \times 1080\) output resolution. Our texture-space approach
only schedules cache updates when cloth elements are being actively deformed and reuses previously
evaluated maps when deformation remains unchanged. We enable all of the optimization techniques discussed
in Section \ref{sec:method_rendering} such as update scheduling, selective
updates to changed tiles, and adaptive virtual textures, we set a max resolution of 4k for the highest quality virtual texture
and we define 5 LODs in total which we swap based on approximate texture texel to viewport pixel footprint.

\begin{figure}[!b]
\captionsetup{skip=9.3pt,belowskip=28.1pt}

\centering
\includegraphics[width=\linewidth]{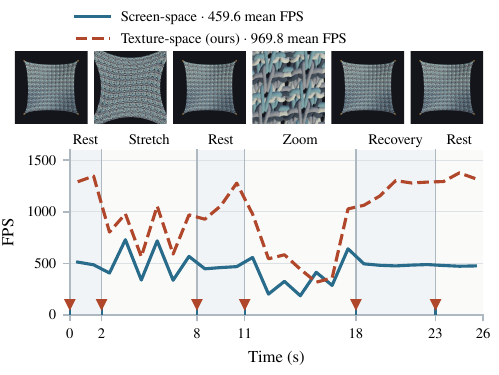}
\caption{
We evaluate the framerate of both per-fragment screen-space deferred rendering and our texture-space approach. We define a scenario where we position a patch of cloth pinned by its four corners looking at the camera. We then deform this patch by stretching and compressing it by its corners, triggering different deformation states from extreme stretch to extreme compression, we have a period of rest, after that we zoom in and then zoom out while the patch is being constantly stretched and compressed, finally we go back to the initial state and we let the garment reach a rest equilibrium.
}
\label{fig:renderer_interaction}
\end{figure}

The benchmark alternates stationary intervals with cloth stretching and a camera zoom
during stretching. We measure GPU-finished application frames per second with browser-driven scheduling 
and vsync disable, we also allow up to three frames in flight.
Figure~\ref{fig:renderer_interaction} reports performance over the interaction sequence.

In this benchmark, we evaluate the performance profile of each method: In the case of the screen-space deferred
approach, we can see that the framerate scales with amount of pixels on screen that require a \methodabbr network evaluation.
Comparatively, in our approach, we can see that we achieve considerably higher framerate in the rest sections of the benchmark, this is because once we calculate that deformation state, we can keep reusing the results and only update the minor changes. We drop in framerate as the fabric starts moving and we require cache updates. We also show the performance behavior of zooming in in both cases,
in screens-pace deferred, performance drops expectedly because we fill the entire viewport with pixels that need network evaluations,
in our approach we also drop in framerate briefly because, as we zoom in, we start swapping to higher quality LODs and calculating them into the cache. As soon as we finish we get a spike again of framerate.

\section{Ablations}
\label{sec:ablations}

We evaluate the main design choices of \methodabbr through ablations of the neural
representation (Section~\ref{ablations:neural_model}) and our texture-cache runtime rendering strategy (Section~\ref{ablations:runtime_rendering}).

\subsection{Neural Model}
\label{ablations:neural_model}

We ablate the architecture, training budget, and frequency initialization of our neural
representation. At approximately matched parameter counts, Fourier-SIREN achieves the
highest reconstruction quality among the evaluated architectures
(Table~\ref{tab:ablation_backbone}). We perform various runs per architecture, sweeping over multiple
learning rates and architecture specific parameters. For our Fourier-SIREN network, increasing the training budget improves
reconstruction quality with diminishing returns beyond 30k steps
(Figure~\ref{fig:ablation_training_length}), which we use throughout our experiments.
(see supplementary material for more details).

\begingroup
\setlength{\tabcolsep}{3pt}
\resulttable{ablation_backbone}{Reconstruction quality of six neural architectures on
\textsc{rib} with approximately 64k parameters each.}
\endgroup

\begin{figure}[!b]
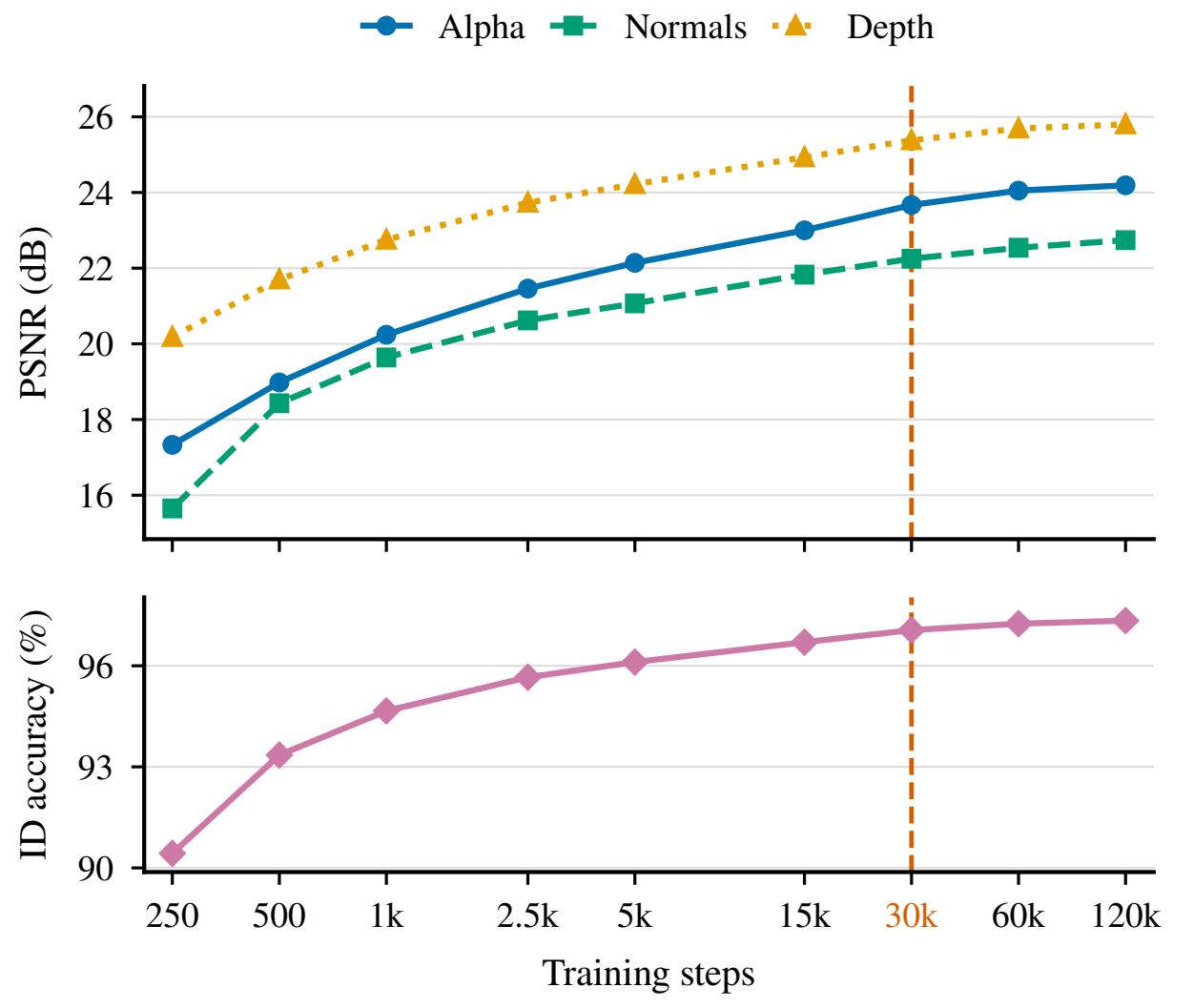

\captionsetup{skip=9.3pt,belowskip=14.6pt}
  \centering
  \genfig[width=\columnwidth]{ablations/ablation_training_length.pdf}
  \caption{Reconstruction quality versus training steps for Fourier-SIREN on
  \textsc{honey}. Top, PSNR (dB) for our intrinsic maps. Bottom, accuracy (\%) for material-ID. The vertical guide marks the 30,000-step budget used in our
  experiments.}
  \label{fig:ablation_training_length}
\end{figure}

\subsection{Runtime Rendering}
\label{ablations:runtime_rendering}

We evaluate our rendering approach along with all the approximations and optimizations discussed in \ref{sec:method} by measuring the end-to-end FPS across multiple scenarios in our Three.js WebGPU renderer implementation. We progressively enable various optimizations and measure the impact that they have on performance. (see the supplementary for the full study)

\section{Limitations}
\label{sec:limitations}

\paragraph*{Source fidelity and material scope.}
DTM inherits the limitations of the microgeometry used for training. Our textile instantiation uses MADYPG centerlines with fixed-radius cylindrical geometry, omitting effects such as cross-sectional flattening, multi-ply rearrangement, and contact-induced compression. The representation itself is
not tied to this geometry model: richer microgeometry \cite{gomeznogales2025unlockingthicknessmodelingcodimensional} could be used provided that deformation-conditioned training maps can be generated. Similarly, although we evaluate DTM only on yarn-based textiles, the same formulation could apply to other materials whose microgeometry can be parameterized by local deformation, such as skin.

\paragraph*{Generalization.}
Each DTM is trained for a single microstructure over a bounded deformation domain. We demonstrate interpolation between sampled deformation states, but not extrapolation beyond this domain or transfer across yarn structures.
Moreover, our deformation descriptor captures local in-plane stretch and shear, but not effects such as bending or localized compression. Future work could explore shared representations conditioned on microstructure and richer
local deformation descriptors.

\paragraph*{Surface representation.}
DTM trades the generality of explicit 3D microgeometry for a compact, surface-attached representation. The predicted opacity, normal, depth, and material-identity maps capture local yarn structure and relief, but do not fully describe the geometry or optical behavior of the underlying microstructure. As a result, the current representation cannot reproduce effects that require full 3D geometry or volumetric light transport, such as complex silhouettes, deep visibility, transmission, and multiple scattering. Our current shading model is also limited to a generic Cook--Torrance BRDF and does not model yarn-direction-dependent reflectance. Finally, scale-dependent filtering remains unresolved: filtering heterogeneous maps requires different treatment of continuous, directional, and categorical
quantities.

\section{Conclusion}
\label{sec:conclusion}

We presented \methodname, a continuous, compact neural representation that
turns dynamic microgeometry rendering into deformation-conditioned texture evaluation.
Our data-generation pipeline maps deformed yarn geometry to a shared texture-space domain
and produces tileable intrinsic maps for learning. We demonstrated \methodabbr on 5 woven
and knitted textiles, capturing changes in yarn arrangement, coverage, and visibility
under stretch and shear.

Our rendering framework evaluates the learned maps on coarse deforming cloth meshes,
avoiding explicit yarn geometry at runtime.
Lighting and material response remain independent of the learned representation,
supporting relighting and runtime yarn-material editing without retraining.
We have shown interactive frame rates under various cloth deformation scenarios along
with multiple optimization methods to achieve higher performance at the cost of minor quality loss
and latency.

While we have demonstrated \methodabbr in the particularly challenging setting of deformable textiles, the underlying representation is not inherently limited to fabrics. Our results highlight the broader potential of deformation-conditioned neural textures as a compact alternative to explicitly representing fine-scale geometry on deforming surfaces. We believe this opens an interesting direction for extending the approach to other materials and structures in which microscopic geometric changes are driven by macroscopic deformation.

\bibliographystyle{eg-alpha-doi}
\begingroup
\catcode`\&=12\relax
\bibliography{bibliography/references}
\endgroup

\end{document}